\documentclass[aps,preprint,amsmath,amssymb]{revtex4}
\usepackage{graphicx}
\usepackage{dcolumn}
\usepackage{bm}
\usepackage{amssymb}
\usepackage{amsmath}

\begin{document}


\title{Characterizing the Hexagonality of Anodic Aluminium Oxide
Nanoporous Arrays}

\author{Luciano da Fontoura Costa} 
\email{luciano@if.sc.usp.br}
 \affiliation{Instituto de F\'{\i}sica de S\~{a}o Carlos, Universidade
 de S\~{a}o Paulo, Av. Trabalhador S\~{a}o Carlense 400, Caixa Postal
 369, CEP 13560-970, S\~{a}o Carlos, S\~ao Paulo, Brazil}

\author{Gonzalo Riveros} 
 \affiliation{Departamento de Qu\'{i}mica, Facultad de Ciencias,
 Universidad de Chile, P.O. Box 653, Santiago de Chile, Chile. }

\author{Humberto G\'{o}mez, Andrea Cortes} 
 \affiliation{Instituto de Qu\'{i}mica, Facultad de Ciencias,
 Universidad Cat\'{o}lica de Valpara\'{i}so, Casilla 4059,
 Valpara\'{i}so, Chile. }

\author{Maxime Gilles, Enrique A. Dalchiele, Ricardo E. Marotti}
 \affiliation{Instituto de F\'{i}sica, Facultad de Ingenier\'{i}a,
 Julio Herrera y Reissig 565, C.C. 30, 11000, Montevideo, Uruguay. }

\date{15th March 2005}

\begin{abstract}

Nanoporous anodized alumina oxide have been used as templates for
obtaining nanomaterials such as nanowires, which exhibit interesting
electronic, magnetic and optical properties.  This article presents
how the regularity of the spatial distribution of pores in such
templates, which affects several of the physical properties of the
obtained nanomaterials, has been objectively quantified.  The method
uses adaptive thresholding, wave propagation image analysis methods,
as well as an hexagonality measurement which was found to be
particularly suitable because of its locality and invariance to
translation, rotation and scaling.  A comparison between commercial
and laboratory-made samples is presented in order to test the method,
resulting higher hexagonality for the latter type of templates.

\end{abstract}


\maketitle

\section{Introduction}

Nanoporous alumina (Al$_{2}$O$_{3}$) obtained by anodizing aluminum
has emerged as an interesting template for
nanomaterials~\cite{Science}, \cite{Gosele}. The resulting porous
anodic aluminum oxide (AAO) has pore diameter that can be varied from
aprox.  10 nm for the smallest pores, to aprox. 300 nm for the largest
pores, depending on the details of the anodizing electrochemical
process (mainly electrolyte and potential).  The depth of the pore
depends basically on the duration of the process, which allows almost
perfectly aligned cylindrical pores with a depth as large as aprox.
100 $\mu$m \cite{AJP}. With this high aspect ratios it is possible to
grow both metallic and semiconductor nanowires \cite{ZhangChemPhys},
\cite{ZnO}, \cite{APA} or nanotubes \cite{Carbonnanotubes} inside the
pores. These nanostructures can have interesting electronic transport
properties \cite{Carbonnanotubes}, \cite{ZhangPRB},
\cite{APLpolycarbonate} and magnetic properties, significantly
different from the same bulk material \cite{ZhengPRB}, \cite {Guo}. In
any case, the optical properties of the AAO\ templates alone have
shown interesting anisotropic \cite{Anisotropy} and luminescent
properties, not yet well understood \cite{JAPLum},
\cite{APLLum}. Altogether with the simple and inexpensive preparation
process, AAO\ appears as an interesting nanostructured material
template with potential for applications in electronic, photonic and
magnetic nanodevices. Moreover, under certain deposition conditions,
the nanopores distribution in the final material is highly ordered
\cite{Science}, \cite{Gosele}, \cite{BaLi}, with an almost hexagonal
distribution. This particular geometry is suitable for many
applications in both quantum electronic effects and photonic
crystals. Indeed, although the properties of the metallo-dielectric
composite has been studied long ago \cite{JAP1978}, their implication
for 2D photonic crystals has been recently placed into evidence
\cite{WideGap}, \cite{ZeroPermitivity}, \cite{Pendry}.

For all the previous many applications of nanoporous AAO the
characterization of both size and distribution of the resulting
material is very important. The usual characterization method of the
hexagonal ordering is made by taking the Fast Fourier Transformation
(FFT) of a 2D image of the surface of the AAO samples \cite{BaLi},
\cite{FFT1}, \cite {FFT2}. However, this method requires a high
orientational order in the whole region under study, and consequently
cannot take appropriate account of the short range hexagonal ordering
of the AAO samples.  The current work describes how physics-based
image analysis methods have been applied in order to obtain the proper
identification of the pores centers and to quantify the spatial
regularity of such spatial structure in terms of an hexagonality
measurement.  This feature is particularly suitable because of its
locality and invariance to translations, rotations, and scaling.  The
application of such a methodology has indicated that alumina arrays
specifically prepared for this work (from now on referred to as
\emph{laboratory-made}) present substantially higher hexagonality than
commonly used commercial arrays~\cite{APA,WideGap}.

\section{Materials and Methods}

\subsection{Sample Preparation and Characterization}

For the present study the AAO arrays have been fabricated by the
following process according to the two step electrochemical
anodization \cite {Science}, \cite{JpnJAP}. First, high purity
(99.95\%) aluminum foils were degreased in hot acetone and after that
ultrasonically cleaned in isopropyl alcohol. Then, the samples were
annealed at 350 $^{o}$C for 1 hour in air.  Subsequently, they were
submitted to a chemical etching in dilute nitric acid and then in 5\%
NaOH at 60 $^{o}$C for 30 s. Afterwards, the aluminum was first
mechanically polished and then electropolished in a 2:2:4 weight
mixture of H$_{2}$SO$_{4}$:H$_{3}$PO$_{4}$:H$_{2}$O. Then, the
aluminum samples was submitted to a first anodization process at 40
V$_{DC}$ in a 0.3 M oxalic acid solution at either 2 $^{o}$C or 20
$^{o}$C, until an alumina layer thickness of ca. 8 $\mu$m was
achieved. The electrolyte was rigorously stirred during anodization.
After that, and only for some samples, this AAO was dissolved away by
immersing the specimens in a mixed solution of 6\%(wt) H$_{3}$PO$_{4}$
and 1.8\%(wt) CrO$_{3}$ at 60 $^{o}$C for 1 hour. Subsequently, the
aluminum sheets were anodized in the same conditions described above
for 6 hours. Consequently, an AAO array with highly ordered pore
arrays were obtained. The specific treatment to which each sample was
submitted is described in Table I.

For comparison, commercial ANOPORE/registered alumina membranes
purchased from the Whatman Company, with 20nm nominal pore size, were
used.  The AAO arrays were imaged by using a scanning electron
microscopy (SEM) JEOL 5900 LV SEM equipment.

\begin{table}
  \vspace{1cm}
  \begin{tabular}{||l|l|l|l|l|l|l|l|r||}  \hline
   Sample & id. & TT & EP & t$_{1}$ & T$_{1}$ ($^{o}$C) & t$_{2}$ & T$_{2}$ ($^{o}$C) & av.hexag.$\pm$ \\ 
    & & & & & & & & st.dev \\ \hline
  commmercial  &  1 & - & - & - & - & - & - & 0.413 $\pm$ 0.085  \\  \cline{2-9}
  membrane     &  2 & - & - & - & - & - & - & 0.375 $\pm$ 0.073  \\  \cline{2-9}
          &  3 & - & - & - & - & - & - & 0.400 $\pm$ 0.077  \\  \cline{2-9}
          &  4 & - & - & - & - & - & - & 0.393 $\pm$ 0.077  \\  \hline
 laboratory  &  5 & No & Yes & 6 hrs. & 2$\pm 1$ & - & - &0.521 $\pm$ 0.123  \\  \cline{2-9}
  made       &  6 & Yes  & No & 6 hrs. & 2$\pm 1$ & - & - & 0.533 $\pm$ 0.121  \\  \cline{2-9}
          &  7 & Yes & Yes & 6 hrs. & 2$\pm 1$ & - & - & 0.472 $\pm$ 0.112  \\  \cline{2-9}
          &  8 & Yes & No & 2 hrs. & 2$\pm 1$ & 6 hrs. & 2$\pm 1$ & 0.532 $\pm$ 0.116  \\  \cline{2-9}
          &  9 & Yes & YEs & 25 min & 20$\pm 1$ & 6 hrs. & 20$\pm 1$ & 0.545 $\pm$ 0.119  \\  \hline
  \end{tabular}
\end{table}

\subsection{Image Analysis Concepts and Methods}

The obtained gray-level images presented intensity fluctuations and
noise which complicated the automated identification of the pores.  In
order to cope with this problem, an adaptive thresholding
method~\cite{Costa_book} was applied which involved the calculation,
for each pixel $p(x,y)$, of the average gray level $\left< p_{R}(x,y)
\right>$ along a circular window with radius of $R$ pixels.  After
subtracting such a value from $p(x,y)$, the result is compared with a
threshold $T$.  In case $\left< p_{10}(x,y) \right> - p(x,y)$ is
smaller than $T$, the pixel at $(x,y)$ is understood to belong to the
core of the pore.  The values of the involved parameters may vary from
image to image.  All cases treated in the current work assumed $R=10$
and varying thresholds.

The adpative thresholding described above produces clusters of pixels
at the core of those pores which have a minimal contrast quality, so
that pores which are faded or which present low contrast are
disregarded.  Because the clusters of points obtained at the center of
each identified pore may appear disconnected, the closing operation
from mathematical morphology~\cite{Costa_book} is applied in order to
ensure that a single connected cluster is obtained at the center of
each pore. 

Once the connected groups were properly obtained, they were labeled
with consecutive integer values, ranging from 1 to $N_p$, where $N_p$
is the number of pores identified in each image.  The center of mass
of each connected group was obtained and used to represent the center
of each pore, which is henceforth called a \emph{seed}.  Constant
speed propagating wavefronts~\cite{Sethian, Costa_book, wavepro} were
then initiated simultaneously at each of such centers, which was
performed by using the distance-based dilations described
in~\cite{Costa_book, wavepro}, so that the shocks between waves
emanating from different seeds corresponded to the boundaries of the
respective Voronoi tessellation.  Such a tessellation has the
interesting property that every point inside each cell is closer to
the respective seed than to any other seed.

For each Voronoi cell, the adjacent cells were obtained by searching
for neighboring pixels with different labels while following the cell
boundary in clockwise fashion.  Such an approach allowed the adjacent
seeds, namely those having a boundary with the reference cell, to be
obtained in suitable order for calculation of the angles.  Figure1(a)
shows a reference seed $i$ (black dot) and four of its adjacent seeds
(white dots).  The successive angles are represented as $\alpha_1$,
$\alpha_2$, \ldots, $\alpha_{N_i}$, where $N_i$ is the number of
neighbors of the reference seed $i$.  The hexagonarity index $h_i$ of
each cell $i$ can now be defined as in Equation~\ref{eq:hex}.  Those
cells which are adjacent with the image borders are excluded from
calculation.

\begin{equation}  \label{eq:hex}
  h_i = \left( {\sum_{k=1}^{N_i} | \alpha_k - \pi/3 |} +1 \right)
  ^{-1}
\end{equation}  

This measurement can be verified to vary from 0 to 1.  Because this
index only takes into account the angles between the reference seed
and those seeds which are adjacent to it, when applied to a single cell
it may produce maximum value of 1 for cells defined by adjacent cells
which are at different distances from the reference seed.  However,
when applied to all cells of a nanoporous arrays, higher average
hexagonality values are obtained only for a nearly hexagonal
structure, assuming smaller values otherwise.  Indeed, hexagonality
densities characterized by all values equal to 1 can only be obtained
for perfectly hexagonal structures.  This is because the presence of
cells with correct angles but varying distances from adjacent seeds
will necessarily imply perturbations to the neighboring cells,
disrupting the respective angles and reducing the hexagonality
indices.

The hexagonality index is invariant to the position, rotation and size
of the image or portions of it.  Moreover, the index value depends
only on the immediate neighborhood of the reference seed and therefore
is not affected by other parts of the image, as would be the case with
global methods such as those based on the Fourier transform.  This
feature allows the proper quantification of the structural uniformity
even in the case where the arrays contain several domains of
hexagonal structures with different preferential orientations.

\section{Results and Discussion}

Figure 2 illustrates the application of the adopted structural
analysis.  The original images of a commercial membrane (a) and a
laboratory-made alumina template (e) samples were processed in order
to obtain the gray-level uniformization and seeds, shown in (b) and
(f).  The Voronoi tessellations obtained from the previous seeds are
shown in (c) and (g).  Note that the different gray-levels correspond
to the distinct labels assigned to each Voronoi cell. The hexagonality
densities of the commercial membranes and laboratory-made templates
are given in (d) and (h), respectively.  It is clear from such
densities that the laboratory-made samples yielded substantially
higher hexagonality values, corroborating the higher spatial
organization indicated by visual inspection of figures (a) and (e).

The average $\pm$ standard deviations of the hexagonalities obtained
for a set of 4 commercial membranes and 5 laboratory-made alumina
samples are given in the last column of Table I, further corroborating
the higher structural uniformity of the laboratory-made nanoporous
templates.  In effect, it is clearly seen that for all except one of
the laboratory-made samples the mean value of the hexagonality defined
by equation (1)\ is greater than 0.5, while it remains below this
number for the commercial membrane. Moreover the highest values are
obtained for the two samples submitted to both anodization process
(samples 8 and 9).  Althought the values are similar for samples 5 and
6 (and quite smaller for sample 7), their deviation is slightly
smaller for samples 8 and 9. This corroborates the contribution of the
second anodization process in increasing the nanoporous ordering [20,
21].

Note that the standard deviation values of the hexagonalities shown in
Table I for the laboratory-made samples are in all cases greater than
the values obtained for the commercial membranes. These results must
be compared with the hexagonalities distributions shown in Figure 2
(d) and (h). For the commercial membrane, Figure 2 (d), the
distribution is almost symmetric, while for the highly-ordered
laboratory-made samples, Figure 2 (h), the distribution is quite
assymmetric. This assymmetry, that may be originated in the self
organization process which leads to the ordering of the nanopore
array, has its maximum in a value greater than its mean value,
implying the increase of the standard deviation for the
laboratory-made samples.

\section{Concluding Remarks}

We have shown how the structural uniformity of nanoporous template,
which have several interesting properties and applications, can be
quantified by using powerful image and shape analysis concepts and
methods, including image correction, mathematical morphology and
Voronoi diagrams.  By taking into account the angular regularity of
individual Voronoi cells, the hexagonality index resulted local and
invariant to position, rotation and scaling of the cells, allowing the
proper quantification of the overall hexagonality even in cases where
the arrays exhibit several domains with different preferential
orientations.  By using such a methodology, it was possible to
conclude that the laboratory-made AAO\ samples present higher
structural uniformity than commercial membranes, and samples with
second anodization process have higher regularity. Future works
include the application of similar analysis to very highly ordered
structures, as well as the extension of the present image treatments
for the determination of other parameters such as pore diameter and
interpore-distance.

\begin{acknowledgments}

LDFC thanks FAPESP (proc. 99/12765-2) and CNPq (proc. 3082231/03-1)
for financial support.  G.R., H.G. and A.C. thank FONDECYT
(Chile). E.A.D. and R.E.M. thank to CSIC (Universidad de la
Rep\'{u}blica) and to PEDECIBA-FISICA, Uruguay.  The authors are also
grateful to J. Troccoli and A. Marquez for SEM measurements.

\end{acknowledgments}

\bibliography{mosaics}

\pagebreak

.

\begin{figure}
\begin{center}
\includegraphics[scale=0.5]{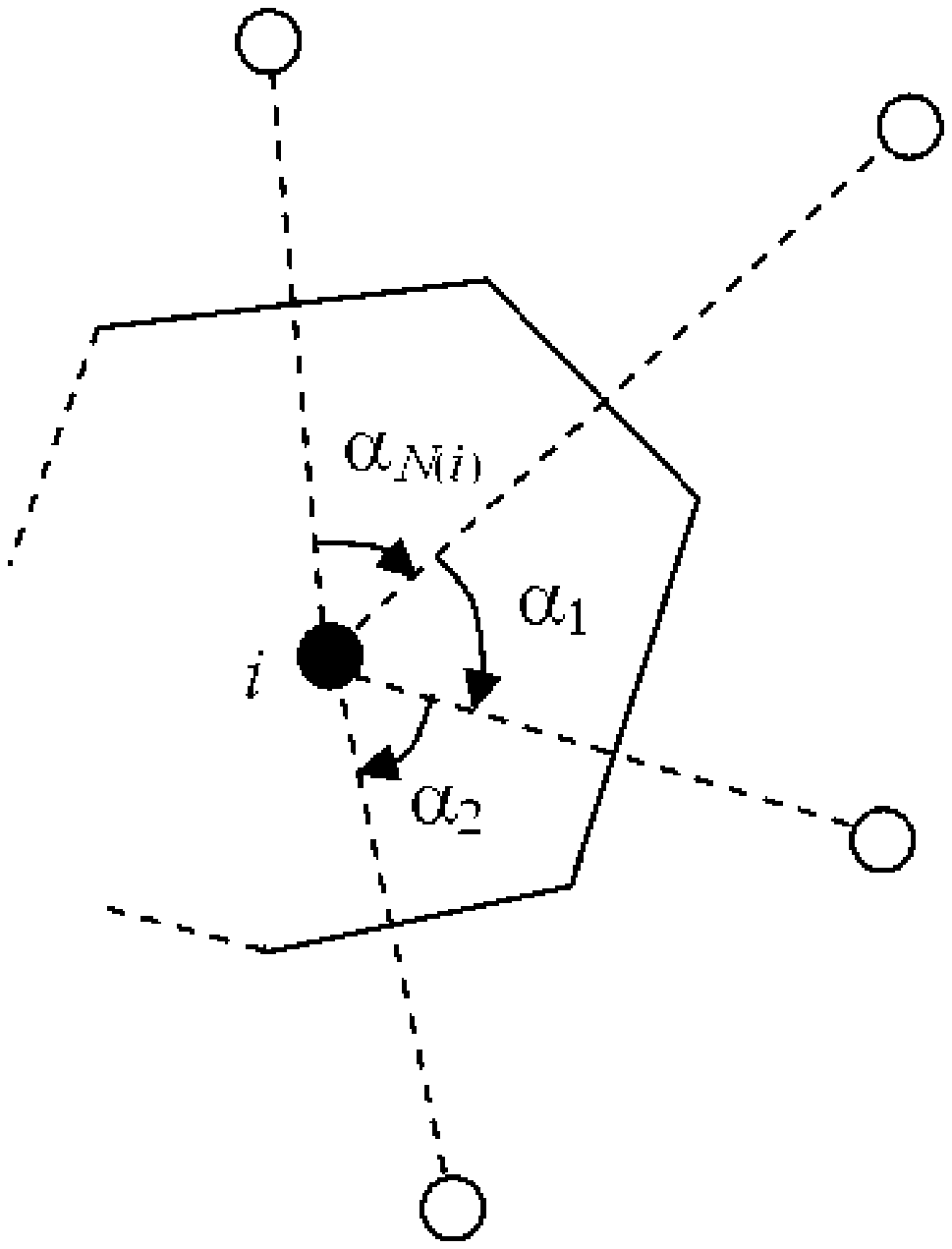} \\
\caption{\label{hexagon} Figure 1.}
\end{center}
\end{figure}

\begin{figure*}
\begin{center}
\includegraphics[scale=0.4212]{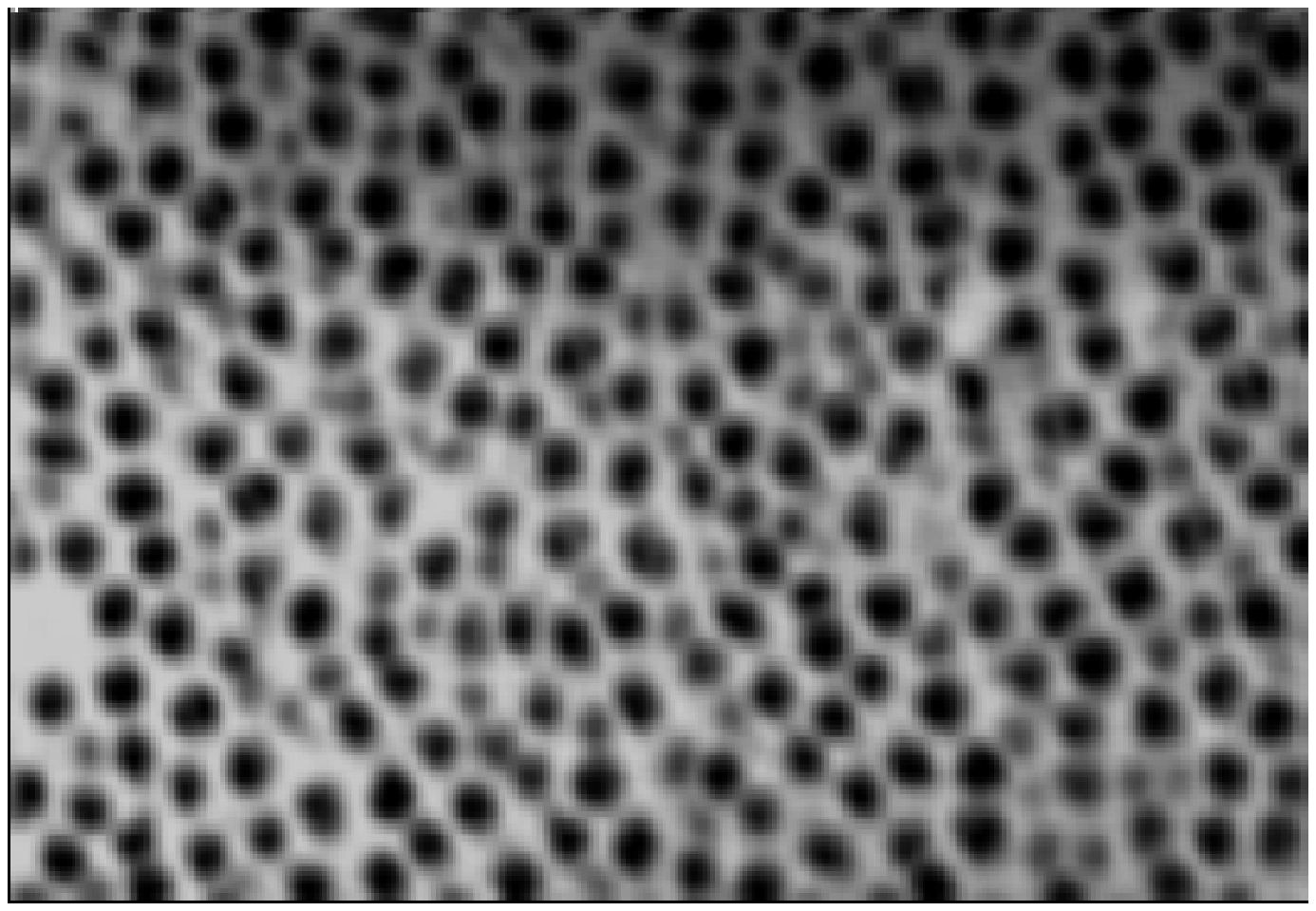} \hspace{0.5cm}
\includegraphics[scale=0.4419]{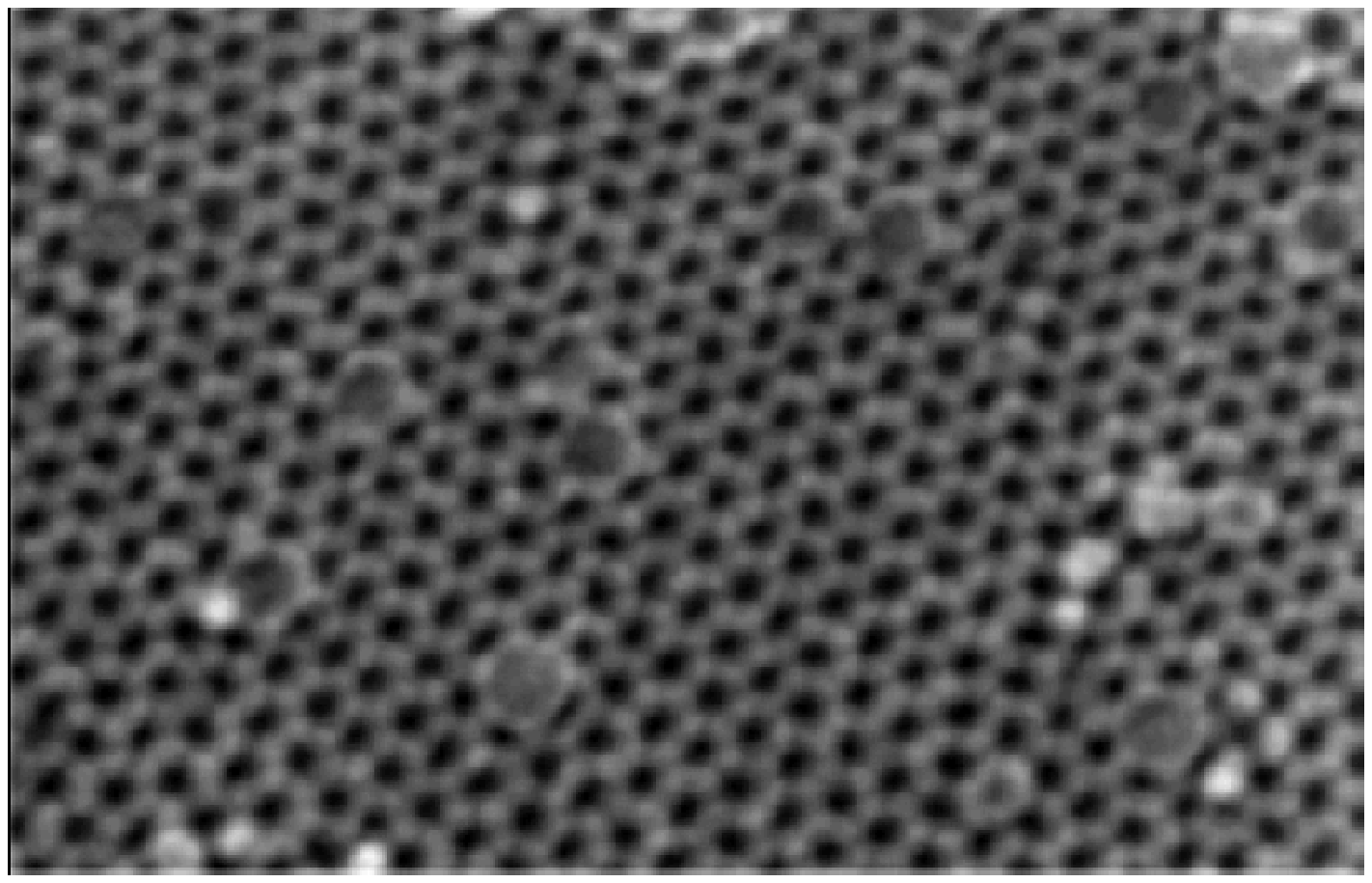} \\
(a)  \hspace{6cm}  (e) \\ \vspace{0.5cm}
\includegraphics[scale=0.405]{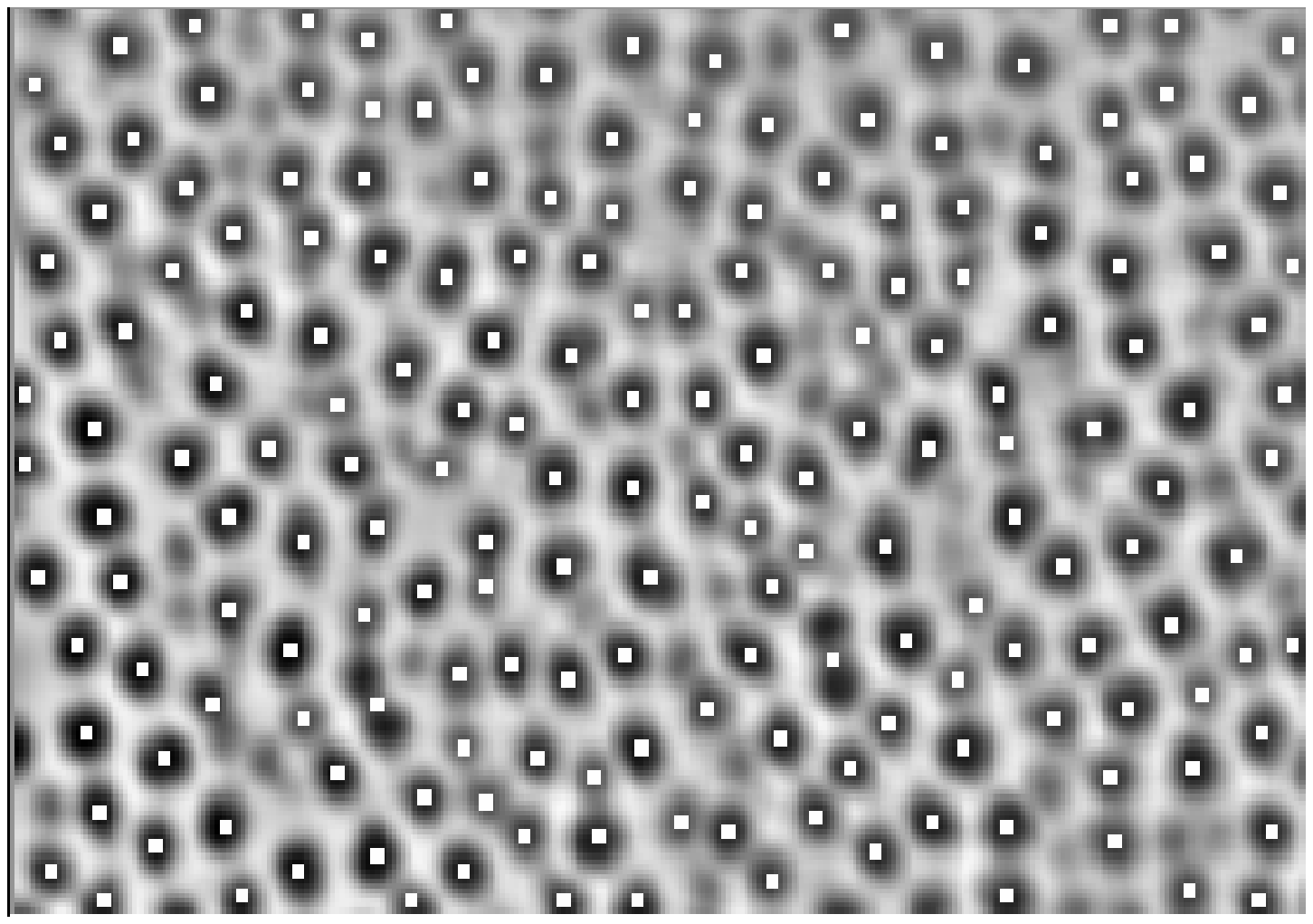} \hspace{0.5cm}
\includegraphics[scale=0.4248]{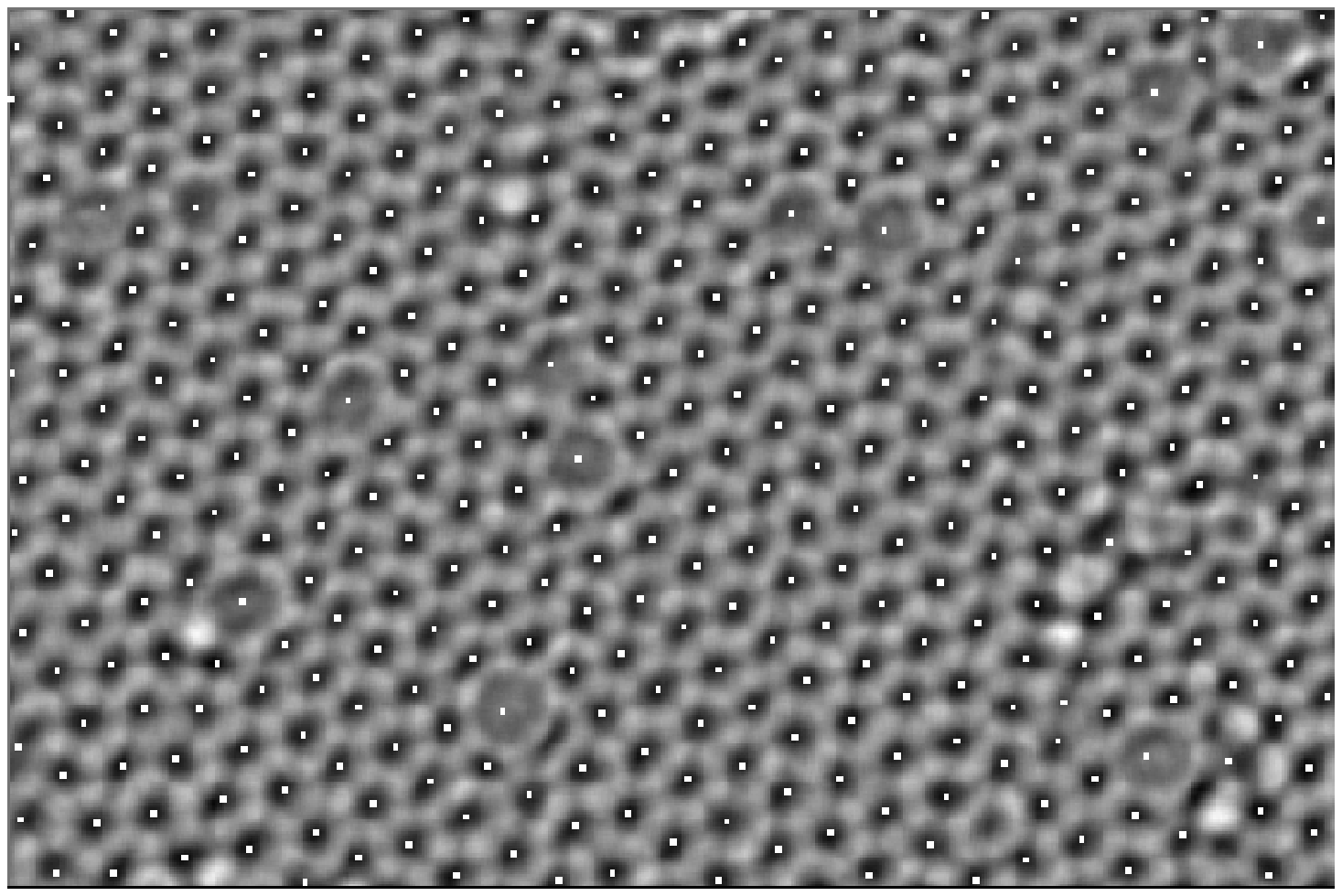} \\
(b)  \hspace{6cm}  (f) \\  \vspace{0.5cm}
\includegraphics[scale=0.477]{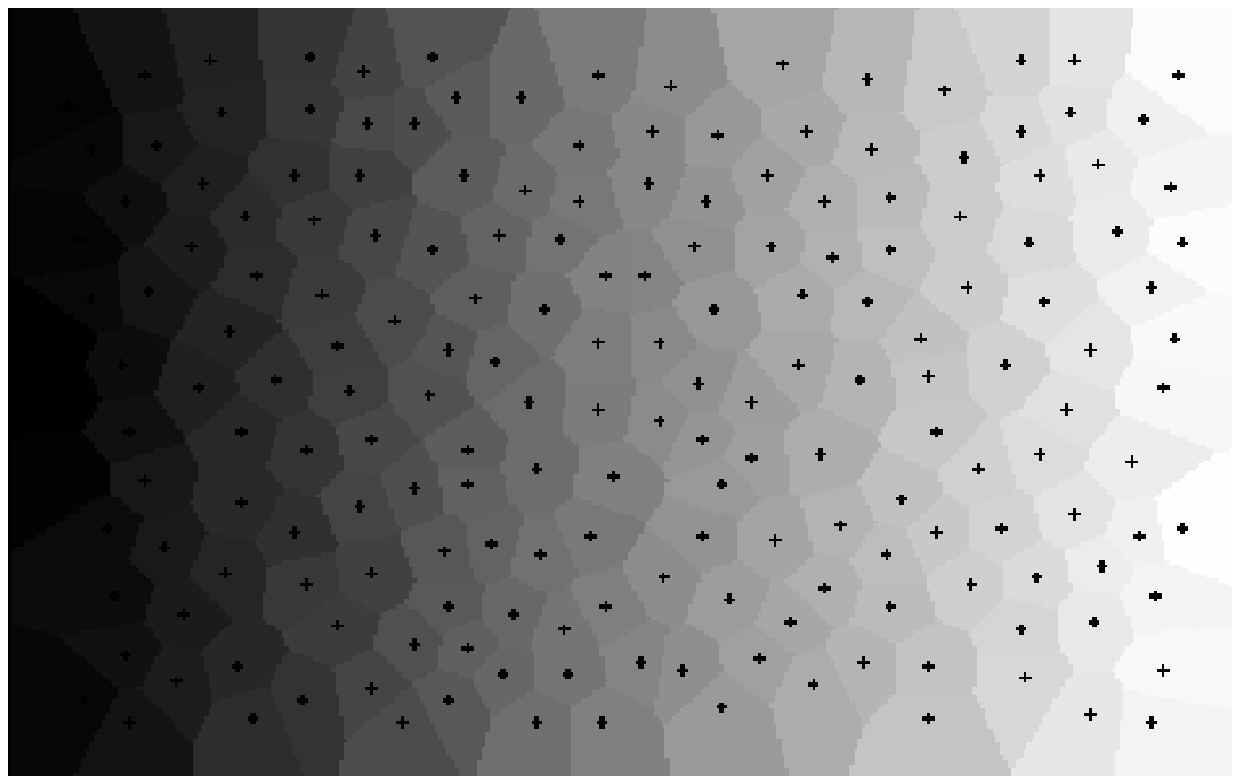} \hspace{0.5cm}
\includegraphics[scale=0.504]{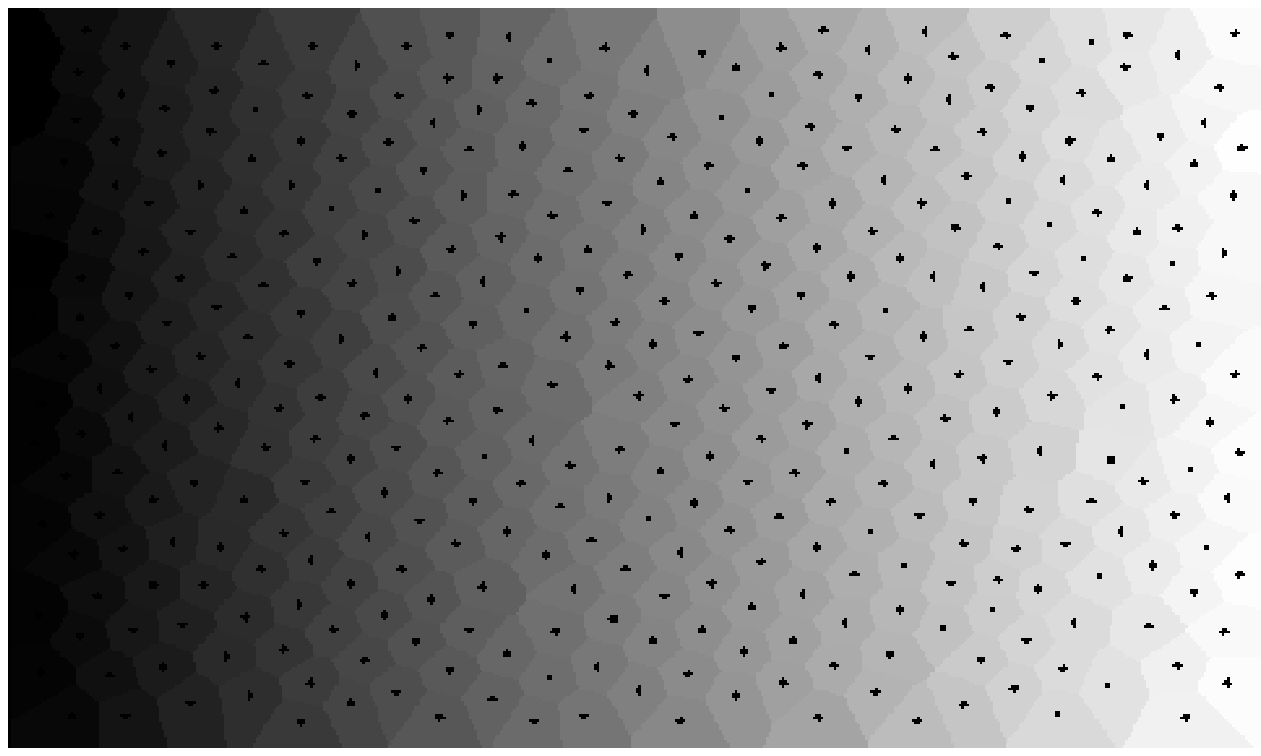} \\
(c)  \hspace{6cm}  (g) \\  \vspace{0.5cm}
\includegraphics[scale=0.405]{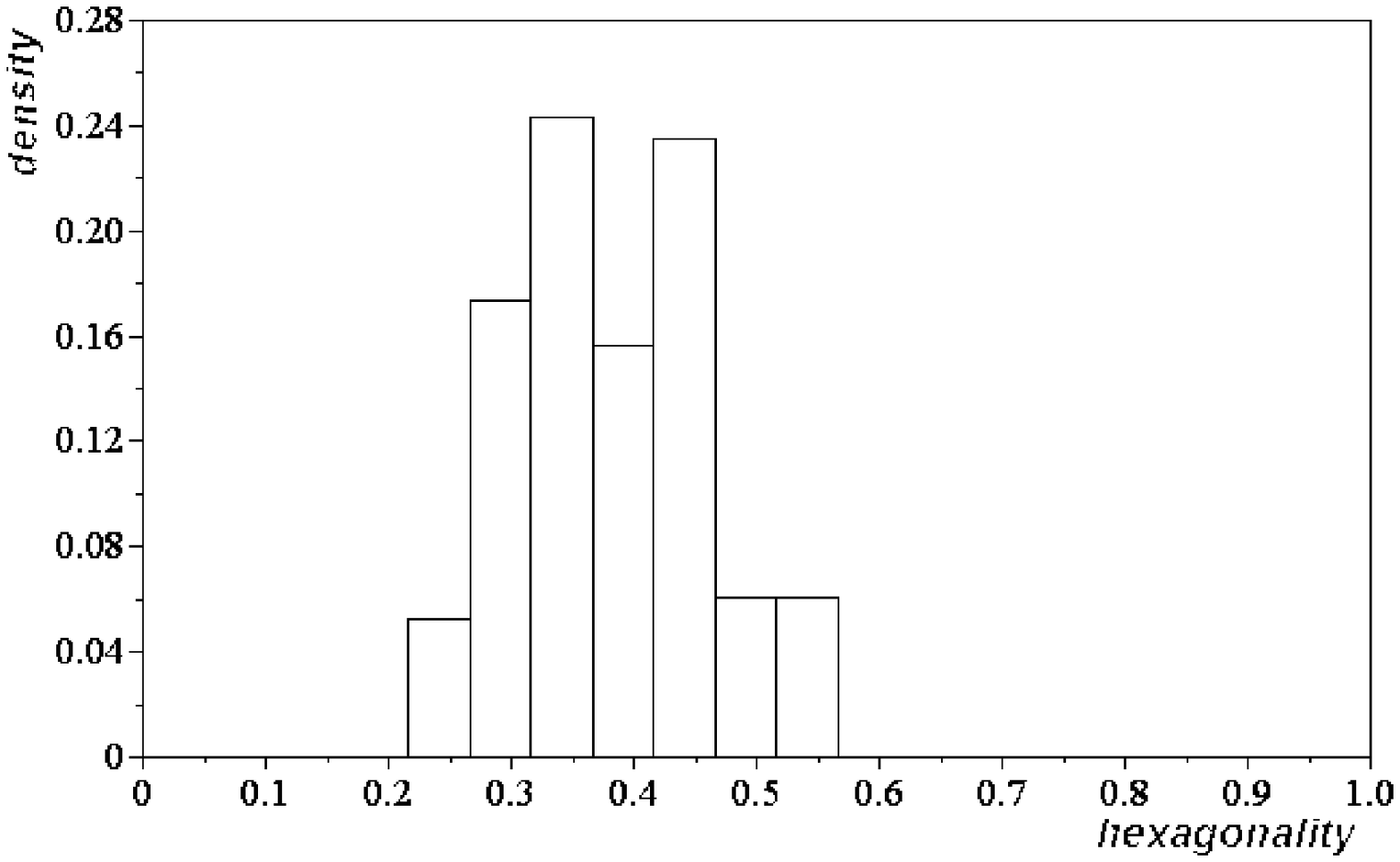} \hspace{0.5cm}
\includegraphics[scale=0.405]{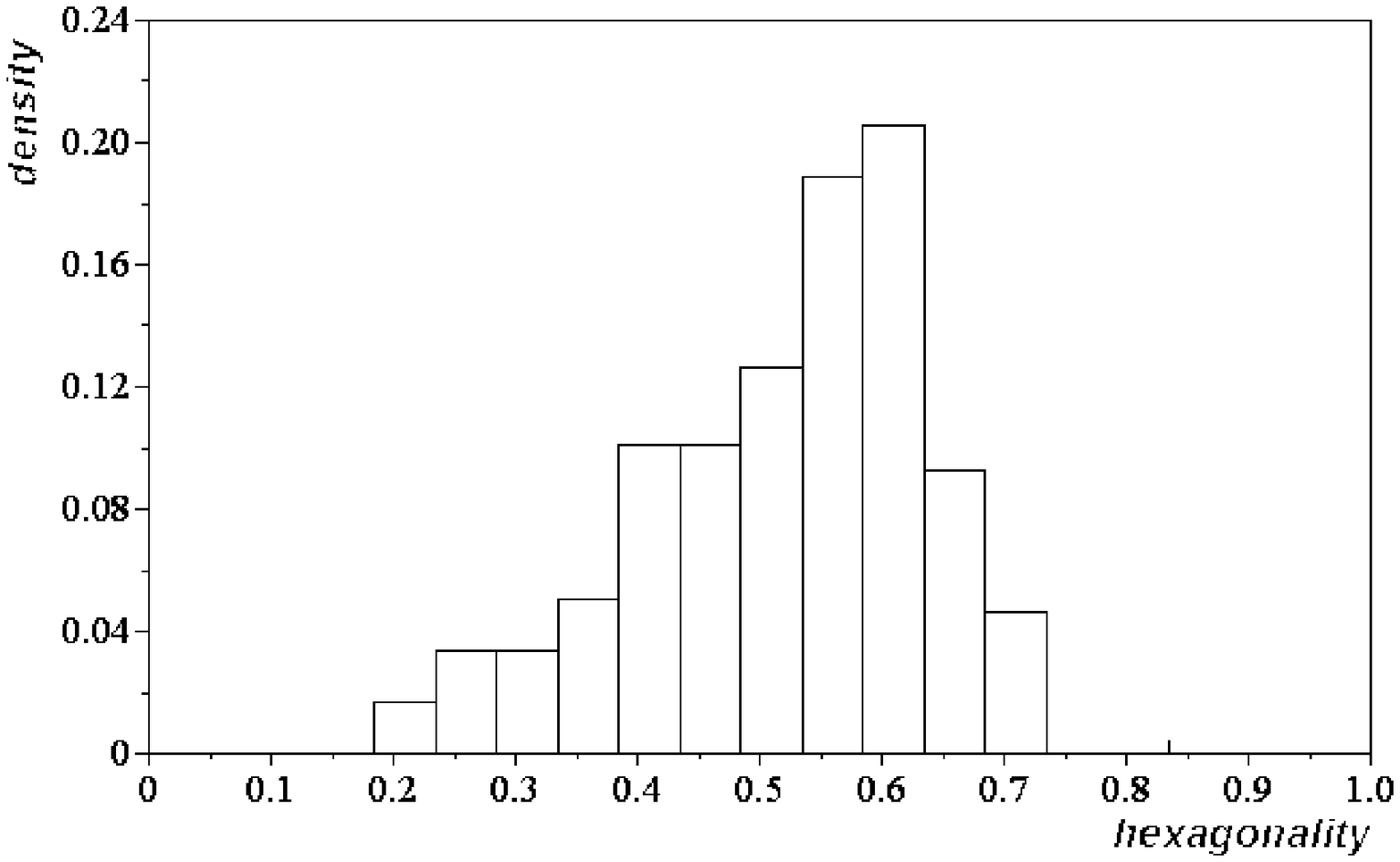} \\
(d)  \hspace{6cm}  (h) \\
\end{center}
\caption{\label{results} Figure 2.}
\end{figure*}

\pagebreak 

{\bf LIST OF CAPTIONS:}

\vspace{1.5cm}

{\bf Figure 1:} Generic Voronoi cell and adopted symbols.

\vspace{1cm}

{\bf Figure 2:} Samples of original commercial (a) and laboratory-made
alumina template (e); respective seeds superimposed onto the
uniformized images (b) and (f); respective Voronoi tessellations (c)
and (g) and densities of the obtained hexagonalities (d) and (h).  It
is clear from the laboratory-made alumina template exhibits
substantially higher hexagonality than the commercial counterpart.

\vspace{1cm}

{\bf Table I:} Description of each sample preparation steps and
respective hexagonalities. TT stands for thermal tretament
(annealing), and EP for electropolishing. T and t corresponds to
temperature and time respectively to the first anodization (subscript
1)\ and second anodization (subscript 2).  Samples 5, 6 and 7 were
submitted to just one anodization process.  The hexagonalities are
given in terms of their averages $\pm$ standard deviations.

\pagebreak

\end{document}